\documentstyle[epsfig,epsf,eqsecnum,aps,amssymb,multicol]{revtex}

\begin{document}
\draft
\preprint{HEP/123-qed}
\title{Hall-conductivity sign change and fluctuations in amorphous
Nb$_{x}$Ge$_{1-x}$ films }
\author{Nobuhito Kokubo, Jan Aarts, and Peter H. Kes \\
}
\address{Kamerlingh Onnes Laboratory, Leiden University,P.O.Box 9504, 2300 RA Leiden, The Netherlands
}

\date{\today}
\maketitle
\begin{abstract}
The sign change in the Hall conductivity has been studied in thin
amorphous Nb$_{1-x}$Ge$_x (x\approx$0.3) films.  By changing the
film thickness it is shown that the field at which the sign
reversal occurs shifts to lower values (from above to below the
mean-field transition field $H_{c2}$) with increasing film
thickness. This effect can be understood in terms of a competition
between a positive normal and a negative fluctuation contribution
to the Hall conductivity.

\end{abstract}
\begin{multicols}{2}
\narrowtext
\section{ INTRODUCTION}
One of the puzzling and intriguing phenomena in type-II
superconductors is the sign change in the Hall effect near the
mean-field transition at the upper critical field $H_{c2}$.  Such
a Hall anomaly has been observed in some conventional low-$T_c$
superconductors, such as moderately disordered Nb and V
\cite{Noto} and amorphous MoSi \cite{Smith,Smith 2} and MoGe
\cite{Graybeal} films, as well as most high-$T_c$ superconductors
(HTSC) \cite{Hagen}.  Hagen et al. \cite{Hagen} pointed out the
importance of the electron mean-free path for the Hall anomaly and
concluded that very clean and very dirty materials do not show
Hall anomalies. However, studies on amorphous dirty
superconductors contradict this conclusion \cite{Smith,Smith
2,Graybeal}.

  Recent phenomenological approaches based
on the time-dependent Ginzburg-Landau (TDGL) equation have
qualitatively explained the sign anomaly \cite{Ikeda ohmi
Tsneto,UD,Kopnin}.  In these theories, the sign reversal is just a
consequence of the difference in sign between the normal (or
quasiparticle) term and the superconducting fluctuation (or vortex
flow) term of the Hall conductivity. Several authors
\cite{FET,Nishio,Aronov,charge} have derived the sign of the
fluctuation (vortex flow) term from the TDGL equation for BCS
superconductors.  Recent experimental studies on HTSCs
\cite{Matsuda2} have pointed out that the sign predictions of
these theories are not correct for HTSCs, but they should be valid
for BCS superconductors.

Even if the sign of the Hall fluctuation conductivity were clear,
its temperature and field dependence is a matter of discussion.
Recent experimental studies on YBa$_2$Cu$_3$O$_{7-\delta}$ films
\cite{Liu,Samo,Lang} and single-crystalline
Bi$_2$Sr$_2$CaCu$_2$O$_{8+\delta}$ and
Bi$_{1.95}$Sr$_{1.65}$La$_{0.4}$CuO$_{6+\delta}$ \cite{Jin} have
observed that the sign change takes place above $H_{c2}$, while
other studies have claimed that the sign anomaly takes place below
$H_{c2}$.  In this problem, the definition of $H_{c2}$ as well as
the temperature and field dependence of the Hall fluctuation
conductivity is very important.

As reported in conventional amorphous films \cite{Berghuis} as
well as HTSCs, the longitudinal conductivity in a perpendicular
magnetic field shows a smooth crossover from the paraconducting
regime to flux flow regime around $H_{c2}$, which is strikingly
different from the picture of the conventional fluctuation theory
in which the conductivity due to the direct fluctuation
contributions of the Aslamazov-Larkin (AL) process diverges at
$H_{c2}$ \cite{Maki}. Thus, it was difficult to define $H_{c2}$
correctly from the fluctuation theory. Recent TDGL theories
\cite{Ikeda ohmi Tsneto}, however, have successfully explained the
smooth crossover around $H_{c2}$ by taking into account the
interaction term of superconducting fluctuations of the AL process
within the Hartree approximation. Later, Ullah and Dorsey (UD)
\cite{UD} developed this further and proposed a scaling theory for
the longitudinal and Hall conductivities.  This scaling approach
is very useful to determine $H_{c2}$ correctly and to describe the
field and temperature dependence of the conductivities.

In this paper, we present measurements and analysis of the
longitudinal and Hall resistivities $\rho_{xx}$ and $\rho_{yx}$
for thin amorphous (a-) Nb$_{1-x}$Ge$_x$ ($x \approx 0.3$) films
($T_c \approx 3K$) according to the TDGL theories. We confirm that
the smooth crossover in the longitudinal conductivity around
$H_{c2}$ is well explained by the UD scaling theory as was found
previously \cite{Theu}, and determine $H_{c2}$.  We then show that
for the thinner films the sign change in the Hall conductivity
takes place above $H_{c2}$. Contrary to results on HTSCs, we show
that the sign of the Hall conductivity is consistent with the TDGL
theory for BCS superconductors.  We discuss the origin of the sign
reversal observed here.

\section{ EXPERIMENTAL}
The films used in this study were deposited by rf sputtering on Si
substrates held at room temperature in a system with a base
pressure of $10^{-6}$mbar, using $10^{-2}$ mbar Ar gas as a
sputtering gas. The thickness as used were 16, 34, 60 and 163nm.
X-ray diffraction showed the films to be amorphous. The average
composition for each film was determined by electron microprobe
analysis.  The distribution in the composition $\delta x$ is less
than $1\%$. The superconducting mean-field transition temperature
in zero field, $T_c$, was determined from the temperature
dependence of resistivity by using the AL fluctuation theory
\cite{AL}. From a previous systematic study on a-Nb$_{1-x}$Ge$_x$
films \cite{Wordenweber}, the distribution of $T_c$ due to $\delta
x$ is estimated to be less than 18mK ($\delta T/T_c \lesssim
6\times 10^{-3}$) around $x=0.3$. Except for the film thickness,
these films have the following identical parameters; the average
composition $x \approx$ 0.3, $T_{c} \approx$ 3K, the normal state
resistivity $\rho_{xx}^n \approx 2.2\mu\Omega$m, $S\equiv
-d(\mu_0H_{c2})/dT|_{T_c} \approx 2$ T/K, the GL coherence length
at $T$=0 $\xi_{GL}(0) \approx 7.3$ nm and the GL parameter for
dirty limit $\kappa \approx $75.  These films were ion-etched in
200$\mu$m wide strips with 8 voltage and 2 current contacts.  The
longitudinal and Hall resistivities are measured by a conventional
dc four-probe method. The longitudinal component due to the
misalignment in the Hall probes was subtracted by reversing the
field direction.  The films are immersed in liquid $^4$He to
obtain good thermal contact. The magnetic field is normal to the
film surface.  The normal resistivity $\rho_{xx}^n$ in the
temperature range of 1.5 K$ < T < $ 5 K has a small temperature
coefficient $(\rho_{xx}^n)^{-1} d\rho_{xx}^n /dT \sim - 10^{-4}$
K$^{-1}$.

\section{ RESULT AND DISCUSSION}
\renewcommand{\theequation}{\arabic{equation}}
 In this study,
$\rho_{xx} (=E_x/J_x)$ and $\rho_{yx} (=E_y/J_x)$ were measured as
a function of $H$ ($|\mu_0H|\leqq$8T) at various $T$.  Figures 1
(a) and (b) show the field dependence of the longitudinal
$\sigma_{xx}(\equiv\rho_{xx}/(\rho_{xx}^2+\rho_{yx}^2))$ and Hall
conductivities
$\sigma_{xy}(\equiv\rho_{yx}/(\rho_{xx}^2+\rho_{yx}^2))$ at
different $T$ for the 34nm thick film with $T_c$=2.77K.  To reduce
the effect of pinning in the mixed state, the measuring current
density $J$ was selected to be 1-4$\times$10$^7$A/m$^2$ which is
much higher than the depinning current density $J_c(\sim 10^5$
A/m$^2$), but smaller than the depairing current density $(\sim
10^{10}$ A/m$^2$).

Far above $T_c$, $\sigma_{xx}$ is field independent while
$\sigma_{xy}$ is directly proportional to $H$, that is, the normal
state Hall effect appears. The normal state Hall conductivity
$\sigma_{xy}^n$ has a positive sign. Within the Drude model, the
normal state Hall angle, $\tan\theta_H^n$, is given by

\begin{equation}
\tan\theta_H^n\equiv\sigma_{xy}^n/\sigma_{xx}^n=\omega_c\tau,
\end{equation}
where $\omega_c$ is the cyclotron frequency and $\tau$ is the
elastic scattering time of electrons. Compared with typical result
on HTSCs ($\omega_c\tau \sim 10^{-2}$ at $\mu_0H=1$T), the present
films have very small value of $\omega_c\tau \sim 10^{-5}$ at
$\mu_0H=1$T, indicating the very small mean-free path to be
expected for amorphous metals.

Near and below $T_c$ one can clearly see that $\sigma_{xy}$
changes sign at a certain field $H^*$ in Fig.1(b). We do not
observe any second sign change below $H^*$, in contrast to what
has been reported for several HTSCs \cite{Kang}. Far above $H^*$,
$\sigma_{xy}$ recovers the direct proportionality to $H$ and the
normal state Hall effect appears again, indicating that the
superconducting fluctuations are completely suppressed by magnetic
field. We therefore can define $\sigma_{xy}^n$ below $T_c$
unambiguously.

In order to determine $H_{c2}$, we use the UD scaling theory.
According to this theory, the longitudinal conductivity is
composed of the normal (or quasiparticle) term $\sigma_{xx}^n$ and
superconducting fluctuation (or vortex flow) term
$\delta\sigma_{xx}$, and expressed as
\begin{equation}
\sigma_{xx}=\sigma_{xx}^n+\delta\sigma_{xx}.
\end{equation}
$\delta\sigma_{xx}$ interpolates smoothly from the paraconducting
regime to flux flow regime around $H_{c2}$ and obeys universal
scaling functions $\widetilde{F}_{\pm}$ where $\widetilde{F}_+
(\widetilde{F}_-)$ is the scaling function for $H>H_{c2} (H
<H_{c2})$.  These functions depend on the dimensionality governed
by the ratio of the film thickness $d$ and the length scale $\xi$
for fluctuations of the order parameter near $H_{c2}$.
For the thickness of the films in this study we can apply
two-dimensional (2D) scaling functions \cite{Theu}.  At each $T$
we identify $\sigma_{xx}^n$ with $\sigma_{xx}$ taken at a field
(typically 7T) where $\sigma_{xy}$ depends linearly on field and
$\sigma_{xx}$ is field independent. $\delta\sigma_{xx}$ is
obtained by subtracting $\sigma_{xx}^n$ from $\sigma_{xx}$. Figure
2 shows a typical scaling result. Here, the data are plotted above
$H_{c2}(T)/3$ where the lowest Landau level (LLL) approximation
for the scaling functions is valid \cite{Theu}. One can clearly
see that the scaled longitudinal fluctuation conductivity
$\widetilde{F}_{xx}^{2D}(\equiv\delta\sigma_{xx}/(C\sigma_{xx}^n(A^{2D}_0t/h)^{1/2}))$
collapses on two universal curves $\widetilde{F}_{\pm}$ as a
function of the scaled field $x^{2D}$ given by $x^{2D}\equiv
\epsilon_H/\sqrt{A^{2D}_0th}$, with
$\epsilon_H=\mu_0(H-H_{c2}(T))/ST_c$, although deviations are
visible at large $|x^{2D}|$.  Here, $t=T/T_c$ and $h=\mu_0H/ST_c$
are normalized temperature and field, respectively.  $C$ is
related to the real part of the relaxation time of the order
parameter $\gamma=\gamma_1+i\gamma_2$.  We take a dirty limit
value of $C=1.447$ \cite{Theu}. The strength of thermal
fluctuations for 2D system, $A^{2D}_0$, is given by
$A^{2D}_0=4\sqrt{2G_i}\xi_{GL}(0)/d$ where $G_i (\approx 5 \times
10^{-6})$ is the Ginzburg number \cite{Lobb}. Very close to
$H_{c2}$, deviations due to the inhomogeneity in the composition
$\delta x$ become apparent. Hence, we do not plot the data in
fields $|\epsilon_H | < (1/2)\delta T_c/T_c\approx 3\times
10^{-3}$, which roughly corresponds to $|x^{2D} |< 0.2$. In such a
scaling plot, the unknown parameters are $S$ and $H_{c2}(T)$. In
the temperature range close to $T_c$, they are connected by the
simple relation $H_{c2}(T)=S(T_c-T)$.  We first determine $S$ from
the scaling collapse of the data close to $T_c$ and this $S$ value
is used to determine $H_{c2}$ far below $T_c$ in the scaling
analysis. Thus, we can unambiguously determine $H_{c2}$ from the
scaling collapse of the data.

  Before proceeding to the result of the $H_{c2}$ line, we
compare the scaling functions $\widetilde{F}_\pm$ with those
predicted in the UD theory.  The UD theory implies that the 2D
universal functions $\widetilde{F}^{2D}_\pm$ in the high field
limit are given by
\begin{equation}
x^{2D}=1/\widetilde{F}^{2D}-\widetilde{F}^{2D},
\end{equation}
if the pinning effect in the flux flow regime is negligible and
the fluctuation conductivity in the paraconducting regime is
dominated by the direct fluctuation contributions of the AL
process.  These functions are applicable to the field range where
the LLL is satisfied.
The solid lines in Fig.2(a) denote these universal functions.
$\widetilde{F}_\pm$ agrees well with $\widetilde{F}^{2D}_\pm$ near
$H_{c2} (-1 \lesssim x_{2D}\lesssim 6$), while deviations are
visible in the large $|x^{2D}|$ regime.
In the paraconducting regime, $\widetilde{F}_+$ decreases much
faster than $\widetilde{F}_+^{2D}$ above $x^{2D} \approx 6$.
Such a rapid decrease in $\delta\sigma_{xx}$ was also observed far
above $H_{c2}$ in amorphous thick films and attributed
qualitatively to a phenomenological short wavelength cutoff in the
fluctuation spectrum \cite{Johnson}.  For the other films ($d$=16
and 60nm) except for the thickest film ($d$=163nm) \cite{Scaling},
similar deviation of $\widetilde{F}_+$ begins to appear at almost
the same value of $x^{2D}\approx 6$, although the physical origin
of the short wavelength cutoff is not clear. The definition of
$\sigma_{xx}^n$ does not affect this behavior because the field at
which $\sigma_{xx}^n$ is defined is much larger than the fields of
interest. Hereafter, we regard $x^{2D}= 6$ as the phenomenological
boundary below which $\delta\sigma_{xx}$ is well described by the
UD scaling theory, and discuss our data below this boundary.

From the scaling collapse of $\delta\sigma_{xx}$ we obtained the
$H_{c2}$ line for films with different thickness.  To compare
those results, we plot the normalized mean-field value of
$\mu_0H_{c2}/ST_c (\equiv h_{c2})$ against normalized temperature
$T/T_c$ for different films in Fig. 3. Good agreement is seen for
$H_{c2}$ of all films. The solid line represents the mean-field
line for the dirty limit in the Werthamer-Helfand-Hohenberg (WHH)
theory, which is given by
\begin{equation} \ln (t) = \Psi (1/2) - \Psi
(1/2+(2/\pi^2)h_{c2}/t)
\end{equation}
where $\Psi$ is the digamma function \cite{Werthamer}.  The
$H_{c2}$ line obtained is well approximated by this relation,
giving experimental support for the validity of the UD scaling
theory.

 Next, we turn to results of the Hall fluctuation
conductivity. In the TDGL theories \cite{UD}, the Hall
conductivity also consists of a normal (or quasiparticle) term and
a superconducting fluctuation (or vortex flow) term,
\begin{equation}
\sigma_{xy}(H,T)=\sigma_{xy}^n(H,T)+\delta\sigma_{xy}(H,T).
\end{equation}
Hence, we subtract $\sigma_{xy}^n(H,T)$ from $\sigma_{xy}(H,T)$,
and plot $\delta\sigma_{xy}(H,T)$ against $H$ in Fig.1(c). The
plot shows that $\delta\sigma_{xy}$ always has a negative sign.
$H_{c2}$ is denoted by the long arrows.  With decreasing $H$ the
magnitude of $\delta\sigma_{xy}$ increases monotonically and grows
as 1/$H$ at low $H(\ll H_{c2})$ (not shown) as the TDGL theories
predict \cite{Dorsey}. Thus, the sign reversal of $\sigma_{xy}$ at
$H^*$ always takes place when $\delta\sigma_{xy}$ and
$\sigma_{xy}^n$ are different in sign. Beforehand it is not clear
whether or not $H^*$ is above $H_{c2}$, because $\sigma_{xy}^n$
and $\delta\sigma_{xy}$ depend in different ways on the electronic
structure of the material. As one can see in Fig.3, in the
thinnest film $H^*$ (denoted as open symbols) is always above
$H_{c2}$ but below the phenomenological boundary where scaling
analysis starts to fail.  It may be worth pointing out that $H^*$
decreases monotonically with rising $T$ and terminates finally at
a certain $T^*$ above $T_{c0}$ in zero field.  With increasing
$d$, $H^*$ moves systematically closer to $H_{c2}$ and it finally
shifts below (but very close to) $H_{c2}$ for the thickest film,
implying that the contribution of the negative $\delta\sigma_{xy}$
to positive $\sigma_{xy}^n$ decreases with increasing $d$.  These
results support the view that enhancing the superconducting
fluctuations by reducing $d$ leads to an increasing negative Hall
conductivity working against positive $\sigma_{xy}^n$, which is
responsible for the sign reversal above $H_{c2}$.

 We now discuss the field and temperature
dependence of $\delta\sigma_{xy}$, in comparison with the UD
scaling theory.  According to this theory, $\delta\sigma_{xx}$ and
$\delta\sigma_{xy}$ have the same field and temperature dependence
and their ratio should be independent of $H$ and $T$.  Note that
$\delta\sigma_{xy}/\delta\sigma_{xx}=-\gamma_2/\gamma_1$, the
ratio of the imaginary and real part of $\gamma$ \cite{UD}. We did
not find scaling of $\delta\sigma_{xy}$.  A recent study on
YBa$_2$Cu$_3$O$_{7-\delta}$ films \cite{Liu} has pointed out that
the failure of the scaling of $\delta\sigma_{xy}$ can be
attributed to the additional contributions of the MT process,
which are not taken into account in the UD theory.  However, the
MT process cannot explain the present result because the strong
pair breaking effect in the amorphous dirty films should lead to a
small contribution \cite{Johnson,Timkham}. As one can see in the
inset of Fig.4(a), contrary to the UD scaling theory,
$-\delta\sigma_{xy}/\delta\sigma_{xx}$ at $H_{c2}$ increases
monotonically with cooling.  Similar temperature dependence of
$-\delta\sigma_{xy}/\delta\sigma_{xx}$ has been reported for
amorphous MoSi films \cite{Smith 2}.  We conclude that the main
reason for the scaling failure is the temperature dependence of
$\gamma_2/\gamma_1$. Further microscopic calculations based on the
BCS theory are required to explain this effect .

The field dependence of $-\delta\sigma_{xy}/\delta\sigma_{xx}$ is
shown in Fig. 4 (a) for two current densities. In the field range
$(-1\lesssim x^{2D} \lesssim 6)$ where $\delta\sigma_{xx}$ follows
the UD scaling theory, $-\delta\sigma_{xy}/\delta\sigma_{xx}$ is
independent of $J$ and depends only weakly on $x^{2D}$.  As one
can see in Fig.4 (b), however, in the same field range both
conductivities change almost one decade in magnitude and their
dependences on $x^{2D}$ look very similar. Hence, we believe that
both $\delta\sigma_{xx}$ and $\delta\sigma_{xy}$ in the
paraconducting regime ($0 \leq x^{2D} \lesssim 6$)) are dominated
by the direct fluctuation contributions of the AL process and thus
the contributions of the AL process are responsible for the sign
change of the Hall conductivity above $H_{c2}$.

Finally, we discuss the origin of the sign in $\sigma_{xy}^n$ and
$\delta\sigma_{xy}$ for our amorphous films.  The sign of
$\sigma_{xy}^n$ depends on the sign of the group velocity
$v(\equiv (1/\hbar)\partial \varepsilon/\partial k$) of electrons
at the Fermi level where $\varepsilon$ is the energy and $k$ is
the wave number. Because of the absence of band structure, the
amorphous materials are generally more free-electron-like than
their crystalline counterparts.  Therefore, the simple amorphous
metals generally have negative $\sigma_{xy}^n$ because of a
positive group velocity ($v \propto k >0$) \cite{Mizutani}. Most
of the amorphous transition metals (TMs), however, have positive
$\sigma_{xy}^n$ \cite{Gallagher}. The origin of this positive
$\sigma_{xy}^n$ has been attributed to the s-d hybridization
interaction in the TM, which leads to a negative group velocity
($\partial \varepsilon/\partial k < 0$) at the Fermi level if the
Fermi energy $\varepsilon_F$ lies within the d-band
\cite{Gallagher,Weir,Weir2}.  The TM-metalloid type amorphous
superconductors NbGe as well as MoGe and MoSi belong to amorphous
TMs and have positive $\sigma_{xy}^n$.

In the TDGL theory based on BCS superconductors by Nishio and
Ebisawa \cite{Nishio}, the sign of $\delta\sigma_{xy}$ is
determined by the electron-hole asymmetry, i.e. by the sign of
$-N'$, where $N'(\equiv
dN(\varepsilon)/d\varepsilon|_{\varepsilon=\varepsilon_F})$ is the
energy derivative of the density of states (DOS) $N(\varepsilon)$
at the Fermi energy.  Numerical calculations of the DOS for e.g.
amorphous Ni \cite{Weir} imply that the total DOS near
$\varepsilon_F$ is dominated by the DOS for the d-band whose
energy dependence is characterized by a peak near the center of
d-band $\varepsilon_{d}$ and roughly approximated by a parabolic
energy dependence with negative curvature, i.e., $N(\varepsilon)
\propto -(\varepsilon-\varepsilon_{d})^2$.  Similar energy
dependences of the total DOS have been commonly observed for
various amorphous TM-metalloid alloys by photoemission experiments
\cite{photoemission}. Because Nb is a less than half filled
4d-band metal, $\varepsilon_F$ lies below the center of the
4d-band $\varepsilon_{4d}$. Thus, a-NbGe films have positive $N'$.
The same argument holds for a-MoGe and MoSi, since Mo is also a
less than half filled 4d-band metal. Thus,
sgn$(\delta\sigma_{xy})$=sgn$(-N') < 0$ in both a-NbGe, a-MoGe and
a-MoSi films \cite{Smith,Smith 2,Graybeal}. These findings give
experimental support for the prediction of the sign of
$\delta\sigma_{xy}$ in the TDGL theory for BCS superconductors.

\section{ SUMMARY}
In summary, we have measured the longitudinal and Hall
resistivities for thin films of the dirty superconductor
a-Nb$_{1-x}$Ge$_x$ ($x \approx 0.3$) near $H_{c2}$.  We confirm
that $\delta\sigma_{xx}$ obeys the 2D scaling functions of the UD
fluctuation theory.  We find a good agreement of the obtained
$H_{c2}$ line with the WHH theory, supporting the scaling
procedure. The failure of the scaling collapse of
$\delta\sigma_{xy}$ is attributed to the temperature dependence of
$\gamma_2/\gamma_1$. The Hall conductivity $\sigma_{xy}$ in
thinner films shows a sign change at a certain $H^*$ which is
above $H_{c2}$ but in the regime where $\sigma_{xx}$ follows the
UD theory.  With increasing film thickness, $H^*$ moves closer to
$H_{c2}$ and it finally shifts below (but close to) $H_{c2}$ for
the thickest film. The negative contribution of the
superconducting fluctuations of the AL process working against
positive $\sigma_{xy}^n$ is responsible for the sign change above
$H_{c2}$. The negative sign of $\delta\sigma_{xy}$ in the present
films is consistent with the electron-hole asymmetry in the
framework of the TDGL theory for BCS superconductors.


 \acknowledgments

We are very grateful to R. Ikeda for useful comments and sending
us his manuscripts.  We would like to thank Y.Matsuda for giving
us a copy of his unpublished work.  We acknowledge the
experimental assistance of R. Besseling, M.B.S.Hesselberth, G.L.E.
van Vliet, R.W.A. Hendrikx and T.J. Gortenmulder. This work is
part of the research program of the "Stichting voor Fundamenteel
Onderzoek der Materie" (FOM), which is financially supported by
NWO. One of the authors (N.K.) is financially supported by JSPS
Postdoctoral Fellowships for Research Abroad.


\begin{figure}
\caption{The field dependence of (a) the longitudinal, (b) Hall
and (c) Hall fluctuation conductivities at different $T$ of
2.08K($\circ$), 2.20K($\square$), 2.47K($\triangle$),
2.60K($\diamondsuit$) and 3.71K($\nabla$) for a 34nm thickness
film.  The short and long arrows denote the sign reversal field
$H^*$ and the mean-field transition field $H_{c2}$, respectively.}
\end{figure}
\begin{figure}
\caption{ The scaled fluctuation conductivity plotted as a
function of $|x^{2D}|$ at different $T$ of 2.08K, 2.20K, 2.47K and
2.60K. The current density $J$ is 2.94$\times 10^{7}$A/m$^2$
except for the curve at 2.08K where $J$=1.47$\times
10^{7}$A/m$^2$. The symbols correspond to those in Fig.1. The
solid curves represent the 2D universal scaling functions
$\widetilde{F}^{2D}_\pm$. $S$ is found to be 2.16 from the scaling
collapse of the data taken at 2.47K and 2.60K close to
$T_c=2.77K$. The $H_{c2}$ values for 2.08K and 2.20K are
determined from the scaling collapse of the data using this $S$
value.}
\end{figure}
\begin{figure}
\caption{ $\mu_0H_{c2}/ST_c$ plotted as a function of $T/T_c$ for
different films of 16nm($\blacksquare$), 34nm($\bullet$),
60nm($\blacktriangle$) and 163nm($\blacklozenge$) thickness.  The
solid curve represents the mean-field line in the dirty limit for
the WHH theory.  The corresponding open symbols show
$\mu_0H^*/ST_c$ for the same films plotted against $T/T_c$. The
dashed and dashed-dotted lines represent the phenomenological
boundaries (given in text) for 34nm and 16nm thickness films,
respectively. For clarity, the boundary for 60nm is not shown. }
\end{figure}
\begin{figure}
\caption{(a) The ratio of the fluctuation conductivities,
$-\delta\sigma_{xy}/\delta\sigma_{xx}$, plotted as a function of
$x^{2D}$ at $T=2.08K$ for the 34nm thickness film with different
$J$ of 1.4kA/cm$^2(\circ$) and 4.4kA/cm$^2(\square)$. (b) The
corresponding longitudinal ($\circ,\square$) and Hall fluctuation
conductivities ($\bullet,\blacksquare$) are also plotted as a
function of $x^{2D}$ with different $J$. Inset in (a) shows the
$T/T_c$ dependence of
$\gamma_2/\gamma_1(=-\delta\sigma_{xy}/\delta\sigma_{xx})$ at
$H_{c2}$ with different thickness of 16nm($\square$),
34nm($\circ$), 60nm($\triangle$) and 163nm($\lozenge$).}
\end{figure}

\newpage
\begin{figure}
\begin{center}
\epsfig{file=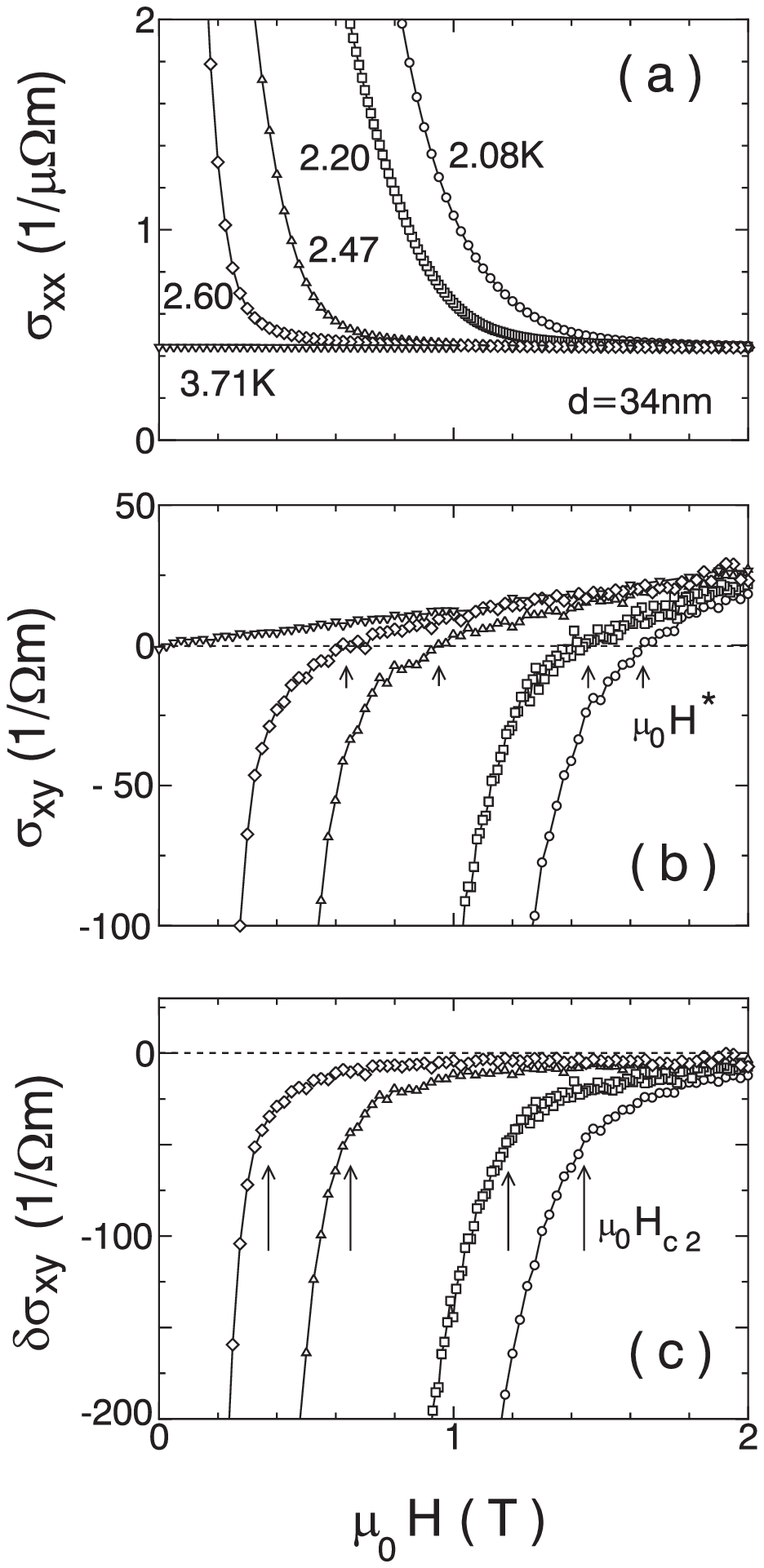,width=5cm} \vspace{0.1cm} Fig.1
\end{center}
\begin{center}
\epsfig{file=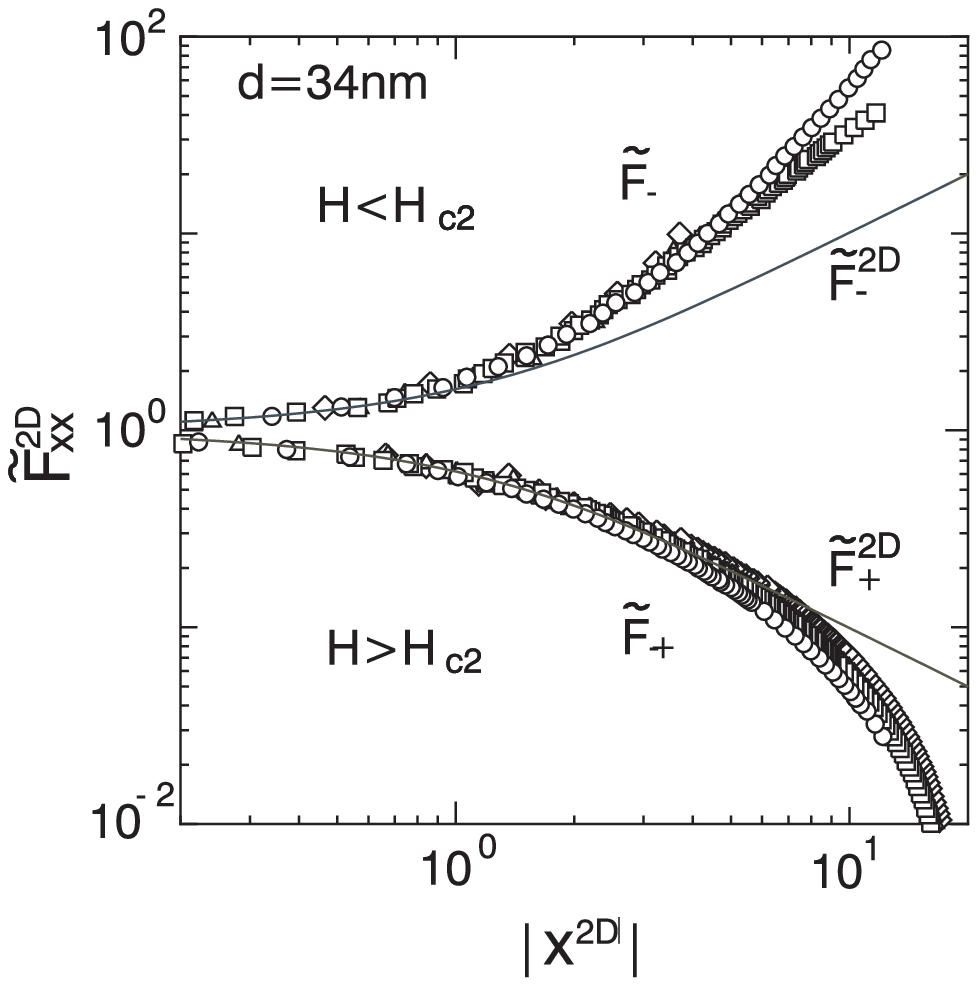,width=5cm} \vspace{0.1cm} Fig.2
\end{center}
\begin{center}
\epsfig{file=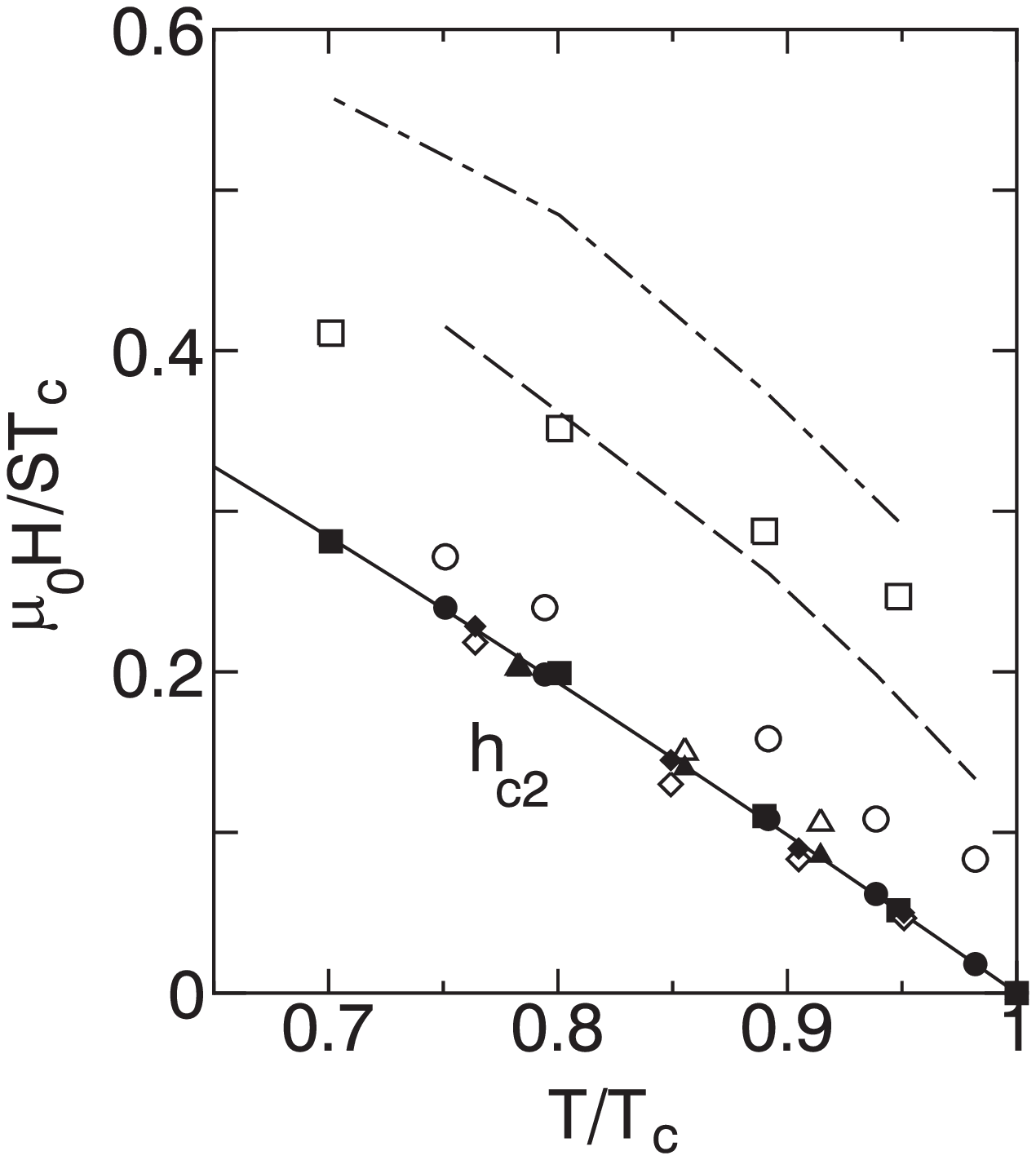,width=5cm} \vspace{0.1cm} Fig.3
\end{center}
\begin{center}
\epsfig{file=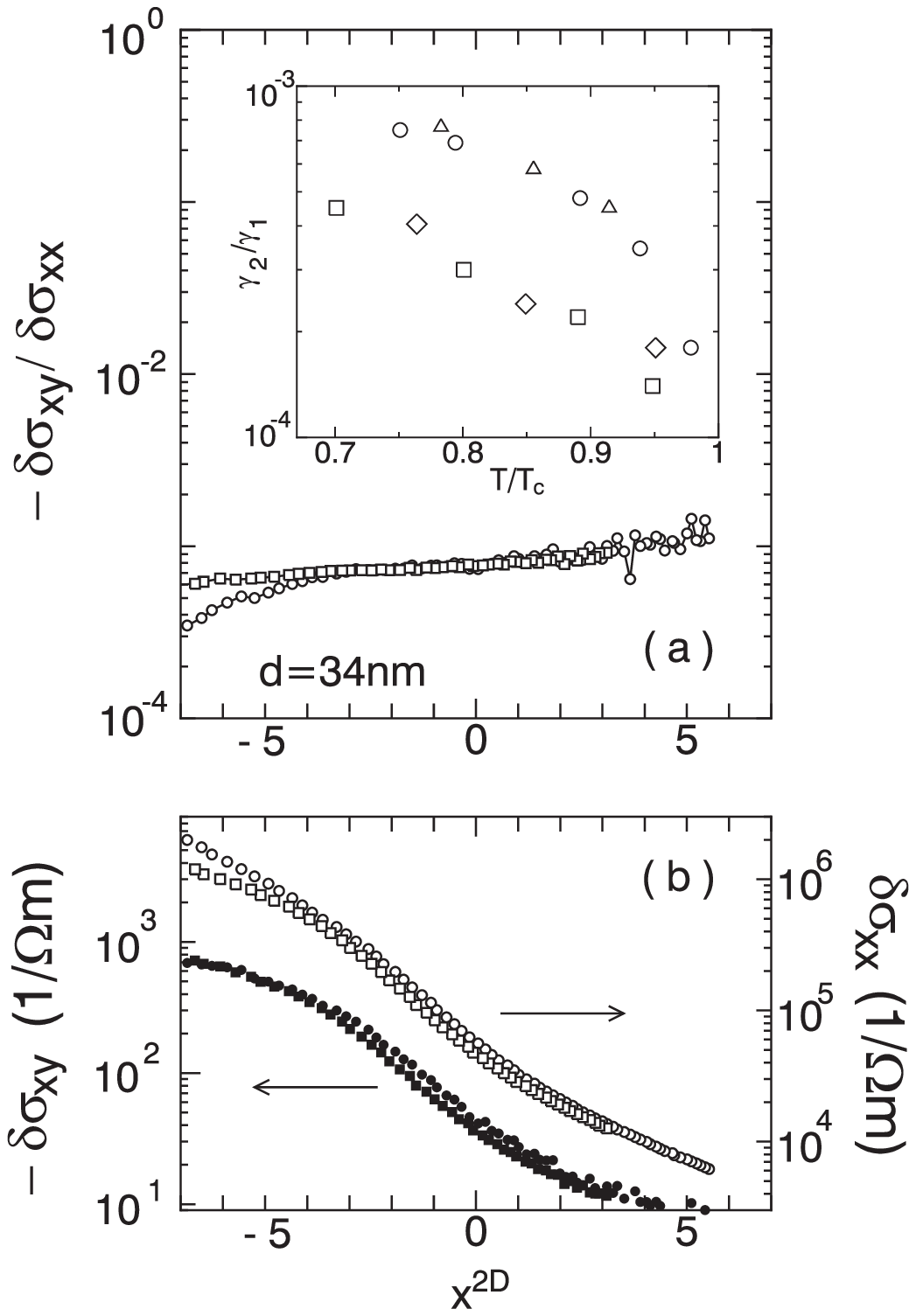,width=5cm} \vspace{0.1cm} Fig.4
\end{center}
\end{figure}
\end{multicols}

\begin{references}
\bibitem{Scaling} The deviations from the 2D scaling functions observed
for the thickest film are different from those observed for the
thinner films.  With increasing $x^{2D}$ $\widetilde{F}_{+}$ for
the thickest film exhibits much slower decrease than
$\widetilde{F}_{+}^{2D}$.  This stems from the dimensional
crossover, although the 3D limit analysis is not fully valid yet
\cite{Theu}. We believe that for the thickest film the 2D scaling
analysis is still valid in a narrow field range near $H_{c2}$.
$H_{c2}$ obtained from the 2D scaling analysis is plotted in
Fig.3.
\end{references}
\end{document}